\documentclass[aps,prl,twocolumn,superscriptaddress,10pt]{revtex4-1}
\usepackage[colorlinks,linkcolor=blue,anchorcolor=blue,citecolor=blue,urlcolor=blue]{hyperref}
\usepackage{graphicx}
\usepackage{epstopdf}
\usepackage{dcolumn}
\usepackage{textcomp}
\usepackage{gensymb}
\usepackage{booktabs}
\usepackage{amsmath}
\usepackage{amssymb}
\usepackage{color}
\usepackage{hyperref}
\usepackage{url}
\usepackage{siunitx}
\usepackage{subfigure}
\usepackage{longtable}
\usepackage{array}
\usepackage{booktabs}
\usepackage{verbatim}
\emergencystretch=\maxdimen
\begin{document}

\title{Positive-parity linear-chain molecular band in \texorpdfstring{$^{16}$C}{}}

\author{\mbox{Y. Liu}}
\author{\mbox{Y.~L. Ye}}
\thanks{yeyl@pku.edu.cn}
\author{\mbox{J.~L. Lou}}
\author{\mbox{X.~F. Yang}}
\affiliation{School of Physics and State Key Laboratory of Nuclear Physics and Technology, Peking University, Beijing 100871, China}
\author{\mbox{T. Baba}}
\affiliation{Kitami Institute of Technology, 090-8507 Kitami, Japan}
\author{\mbox{M. Kimura}}
\affiliation{Department of Physics, Hokkaido University, 060-0810 Sapporo, Japan}
\author{\mbox{B. Yang}}
\author{\mbox{Z.~H. Li}}
\author{\mbox{Q.~T. Li}}
\author{\mbox{J.~Y. Xu}}
\author{\mbox{Y.~C. Ge}}
\author{\mbox{H. Hua}}
\affiliation{School of Physics and State Key Laboratory of Nuclear Physics and Technology, Peking University, Beijing 100871, China}
\author{\mbox{J.~S. Wang}}
\affiliation{School of Science, Huzhou University, Huzhou 313000, China}
\affiliation{Institute of Modern Physics, Chinese Academy of Science, Lanzhou 730000, China}
\author{\mbox{Y.~Y. Yang}}
\author{\mbox{P. Ma}}
\author{\mbox{Z. Bai}}
\author{\mbox{Q. Hu}}
\affiliation{Institute of Modern Physics, Chinese Academy of Science, Lanzhou 730000, China}
\author{\mbox{W. Liu}}
\author{\mbox{K. Ma}}
\author{\mbox{L.~C. Tao}}
\author{\mbox{Y. Jiang}}
\affiliation{School of Physics and State Key Laboratory of Nuclear Physics and Technology, Peking University, Beijing 100871, China}
\author{\mbox{L.~Y. Hu}}
\affiliation{Fundamental Science on Nuclear Safety and Simulation Technology Laboratory, Harbin Engineering University, Harbin, 150001, China}
\author{\mbox{H.~L. Zang}}
\author{\mbox{J. Feng}}
\author{\mbox{H.~Y. Wu}}
\author{\mbox{J.~X. Han}}
\author{\mbox{S.~W. Bai}}
\author{\mbox{G. Li}}
\author{\mbox{H.~Z. Yu}}
\author{\mbox{S.~W. Huang}}
\author{\mbox{Z.~Q. Chen}}
\author{\mbox{X.~H. Sun}}
\author{\mbox{J.~J. Li}}
\author{\mbox{Z.~W. Tan}}
\affiliation{School of Physics and State Key Laboratory of Nuclear Physics and Technology, Peking University, Beijing 100871, China}
\author{\mbox{Z.~H. Gao}}
\author{\mbox{F.~F. Duan}}
\affiliation{Institute of Modern Physics, Chinese Academy of Science, Lanzhou 730000, China}
\author{\mbox{J.~H. Tan}}
\author{\mbox{S.~Q. Sun}}
\author{\mbox{Y.~S. Song}}
\affiliation{Fundamental Science on Nuclear Safety and Simulation Technology Laboratory, Harbin Engineering University, Harbin, 150001, China}

\date{\today}

\begin{abstract}
  An inelastic excitation and cluster-decay experiment $\rm {^2H}(^{16}C,~{^{4}He}+{^{12}Be}~or~{^{6}He}+{^{10}Be}){^2H}$ was carried out to investigate the linear-chain clustering structure in neutron-rich $\rm {^{16}C}$. For the first time, decay-paths from the $\rm {^{16}C}$ resonances to various states of the final nuclei were determined, thanks to the well-resolved $Q$-value spectra obtained from the three-fold coincident measurement. The close-threshold resonance at 16.5~\si{MeV} is assigned as the ${J^\pi}={0^+}$ band head of the predicted positive-parity linear-chain molecular band with ${(3/2_\pi^-)^2}{(1/2_\sigma^-)^2}$ configuration, according to the associated angular correlation and decay analysis. Other members of this band were found at 17.3, 19.4, and 21.6~\si{MeV} based on their selective decay properties, being consistent with the theoretical predictions. Another intriguing high-lying state was observed at 27.2~\si{MeV} which decays almost exclusively to $\rm {^{6}He}+{^{10}Be{(\sim6~\si{MeV})}}$ final channel, corresponding well to another predicted linear-chain structure with the pure $\sigma$-bond configuration.
\end{abstract}

\maketitle
Clustering is a general phenomenon appearing at every hierarchical layer of the matter universe, including the largest star systems~\cite{Star2010} and the smallest hadron systems~\cite{Quark2016}. In light nuclei, cluster formation has been widely adopted to interpret some peculiar occurrences of quantum states together with their particular population and decay properties~\cite{Ikeda1968, Mg24Chain1992, vonOertzen2006, ClusterPPNP2015, FreerCluster2018, ClusterPPNP2020}. In recent years, clustering phenomenon has attracted further attention in the study of unstable nuclei in which the extra valence nucleons may act as covalent bonds to stabilize the whole system~\cite{HoriuchiCluster2012}, analogous to those in atomic molecules~\cite{vonOertzen2006, ClusterPPNP2015, FreerCluster2018, ClusterPPNP2020}. In these studies, cluster-decay measurement has played an essential role. It provides the high sensitivity to the clustering states having much lower level-density, the advantage to determine spin of the resonance from model-independent angular correlation analysis~\cite{Freer1996}, and the possibility to connect the unknown structures of the mother nucleus to the known structures of the detected daughter fragments~\cite{LIJING2017}.

For neutron-rich beryllium isotopes, molecular structures built on the dual-$\alpha$ cores have been extensively studied by configuring the valence neutrons into $\pi$-type or $\sigma$-type covalent bonds~\cite{vonOertzen2006, ClusterPPNP2015, FreerCluster2018}. Similar studies have been naturally extended to the triple-$\alpha$ systems, the carbon isotopes, where the triangle and linear-chain configurations are anticipated~\cite{Morinaga1956, vonOertzen2006}. In recent years, substantial works have been devoted to investigating the linear-chain configurations in $\rm {^{13-14}C}$ and some evidences have been reported in the literature~\cite{C13PRL2019, Feng2019, Freer2014, Fritsch2016, C14PRC2016, LIJING2017, C14PLB2017, C14CPC2018}. Latest antisymmetrized molecular dynamics (AMD) calculations, without using the predefined cluster degrees of freedom, have also predicted several linear-chain molecular bands in $\rm {^{16}C}$~\cite{C16AMD2014, C16Kimura2018}. Most importantly, the calculations have proposed a characteristic decay pattern which collects the members of the positive-parity linear-chain band to the $\rm {^{4}He}$ + $\rm {^{12}Be}$ and $\rm {^{6}He}$ + $\rm {^{10}Be}$ final channels, with the Be fragments at various low-lying states. However, observation of this pattern requires precise measurements allowing to discriminate states in the final nuclei. This relies quite often on the resolution of the reaction $Q$-value.  Unfortunately, so far the experiments aiming at $^{16}$C-clustering have not been able to achieve this requirement due basically to the limited beam quality, detection system performances and statistics~\cite{C16JPG2001, C16PRC2002, C16PRC2004, C16PRC2016}.

In this letter, we report on a new inelastic scattering and cluster-decay experiment for $\rm ^{16}C$, in which all final particles were coincidentally detected with high efficiency. This kind of full particle-detection method has been applied previously to suppress the reaction background in the experiment using the stable nucleus beam ~\cite{Aquila2017}. In our case, this method is essential to deal with the large energy-spread problem resulted from the secondary radioactive ion beam. The beam energy can actually be deduced, event by event, from the three final particles according to the energy-momentum conservation. As a result, the obtained $Q$-value resolution does not rely on the original beam energy spread and allows to reconstruct $^{16}$C excitation spectra based on their decay paths. The predicted positive-parity linear-chain molecular band has been systematically analyzed and confirmed. Another exotic state at 27.2~\si{MeV} was also found to decay primarily into $\rm {^{10}Be(\sim6~\si{MeV})}$, in line with the property of the predicted pure $\sigma$-bond linear-chain band at even higher energies.


The experiment was performed at the Radioactive Ion Beam Line at the Heavy Ion Research Facility in Lanzhou (HIRFL-RIBLL)~\cite{SUN2003}. A 23.5~\si{MeV/nucleon} $^{16}$C secondary beam, with an intensity of about $1.5\times10^4$ particles per second and a purity of about 90\%, was produced from a 59.6~\si{MeV/nucleon} $^{18}$O primary beam impinging on a 4.5~\si{mm} thick $^9$Be target. Three $x$-$y$ position-sensitive parallel plate avalanche chambers were employed to track the $^{16}$C beam onto a 9.53~\si{mg/cm^2} $({\rm CD}_2)_n$ target foils. The deuterium target was chosen owing to its easiness to be detected as a recoil particle and its power to excite the projectile.

\begin{figure}
  \begin{center}
  \includegraphics[width=.40\textwidth]{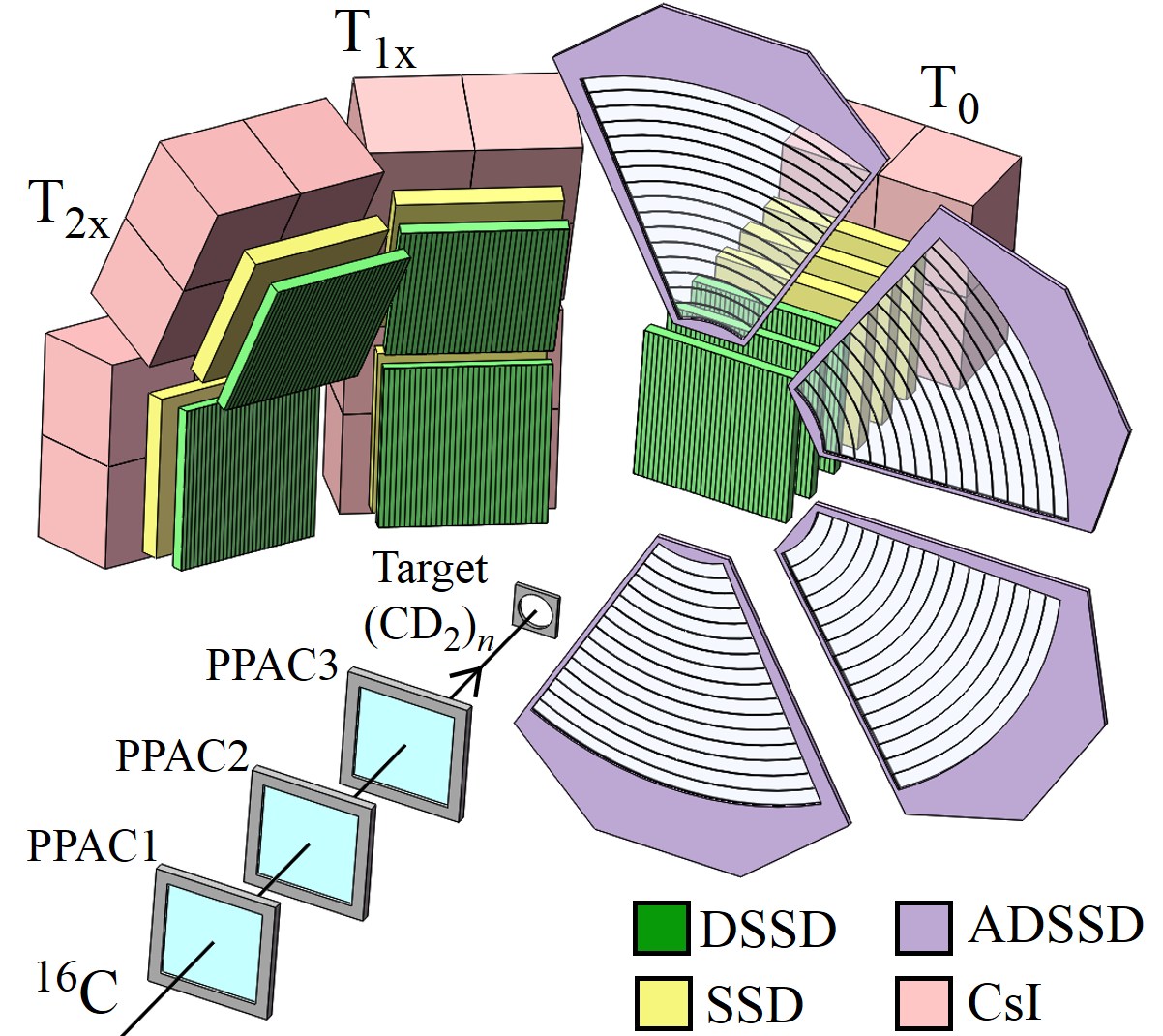}
  \caption{A schematic view of the experimental setup. x in $\rm T_{\rm 1x}$ and $\rm T_{\rm 2x}$ stands for up and down.}
  \vspace{-6mm}
  \label{fig:Setup}
  \end{center}
\end{figure}

A schematic layout of the detection system is given in Fig.~\ref{fig:Setup}. The decaying helium and beryllium fragments, from the $\rm {^2H}(^{16}C, {^4He}+{\!^{12}Be}){^2H}$ and $\rm {^2H}(^{16}C, {^6He}+{\!^{10}Be}){^2H}$ reactions, were coincidentally detected by a zero-degree Si-CsI telescope (T$_0$), while the recoil $^2$H was measured by the annular double-sided silicon strip detectors (ADSSD) and four other Si-CsI telescopes (T$_{\rm 1x}$ and T$_{\rm 2x}$). The T$_0$ telescope was composed of three double-sided silicon strip detectors (DSSD), three single-sided silicon detectors (SSD), and a $2\times2$ CsI(Tl) scintillator array. Each DSSD has a nominal thickness of 1000~{\textmu}m and an active area of $64\times64$~\si{mm^2} with 32 strips on each side of the silicon layer. Each SSD has the same active size as the DSSD while its nominal thickness is 1500~{\textmu}m. The first layer of the T$_0$ array was placed at 156~\si{mm} from the target, accepting almost $100\%$ of the decaying fragments because of the inverse kinematics~\cite{YANG2015, FREER2001}.  The T$_{\rm 1x}$ and T$_{\rm 2x}$ telescopes were centered at 35{\degree} and 69{\degree} with respect to the beam direction, and at distances of  178.7~\si{mm} and 156.6~\si{mm} from the target, respectively. Each of them was composed of a thin DSSD (60 or 300 ~{\textmu}m), a thick SSD (1500~{\textmu}m) and a $2\times2$ CsI(Tl) scintillator array. Four sectors of ADSSD (150 or 400 ~{\textmu}m thick) were installed around T$_0$ telescope at a distance of 123~\si{mm} from the target.

Energy calibration of the detectors was accomplished by using $\alpha$-particle sources and the procedures described in Refs.~\cite{Jenny2018, QIAO2014}. Timing information obtained from the DSSD strips was applied to assure the real coincidence among the recorded signals. This is particularly important for T$_0$ telescope which was directly exposed to the beam.  Particles produced from the reactions on the detector layers, instead of those on the target, were excluded by employing the tracking method. Fake-coincident signals resulted from the inter-strip gap-hitting were also discriminated by matching the tracks and energies in neighboring detector layers. Thanks to the excellent energy, timing and position resolutions of the silicon detectors, isotopes from hydrogen to carbon were unambiguously identified based on the standard energy loss versus residual energy (${\Delta}E$-$E$) technique~\cite{YangPRC2019}. The detection and calibration were validated by using the two- and three-$\alpha$ coincident events to reconstruct the known $^8$Be and $^{12}$C resonances, respectively~\cite{Mg24Chain1992, C16PRC2016, C12Hoyle2017}.

\begin{figure}[htp!]
  \begin{center}
  \subfigure{\label{fig:SpectrumE1}}
  \vspace{-0.1in}
  \hspace{-0.1in}
  \subfigure{\label{fig:SpectrumE2}}
  \vspace{-0.1in}
  \hspace{-0.1in}
  \subfigure{\label{fig:SpectrumQ1}}
  \vspace{-0.1in}
  \hspace{-0.1in}
  \subfigure{\label{fig:SpectrumE3}}
  \vspace{-0.1in}
  \hspace{-0.1in}
  \subfigure{\label{fig:SpectrumE4}}
  \vspace{-0.1in}
  \hspace{-0.1in}
  \subfigure{\label{fig:SpectrumE5}}
  \vspace{-0.1in}
  \hspace{-0.1in}
  \subfigure{\label{fig:SpectrumQ2}}
  \includegraphics[width=.48\textwidth]{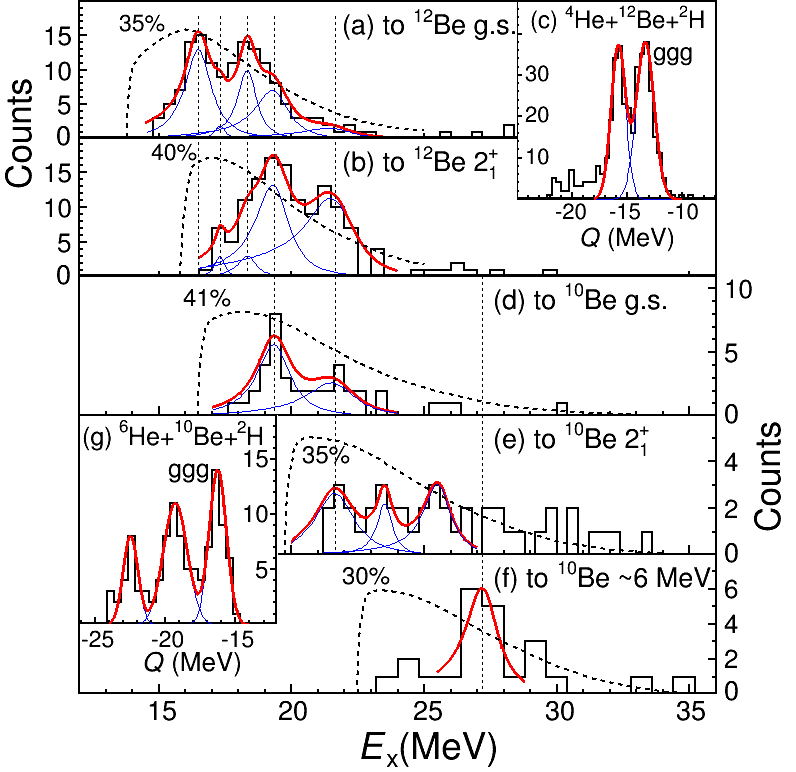}
  \vspace{0.6in}
  \hspace{0.6in}
  \vspace{-5mm}
  \caption{Excitation energy spectra of $\rm {^{16}C}$ reconstructed from two final channels ($^4$He + $^{12}$Be + $^2$H and $^6$He + $^{10}$Be + $^2$H ) and gated on $Q$-value peaks for decaying into $\rm {^{12}Be}$($0_1^+$) (a), $\rm {^{12}Be}$($2_1^+$) (b),$\rm {^{10}Be}$($0_1^+$) (d), $\rm {^{10}Be}$($2_1^+$) (e), and $\rm {^{10}Be}$($\sim6$~\si{MeV}) (f). Each spectrum is fitted by the sum (red-solid line) of several resonant peaks (blue-solid line). The black-dashed lines stand for the detection efficiencies as a function of excitation energy, for each of which the maximum is indicated by the percentage value. The vertical black-dotted lines are plotted to guide the eyes for the corresponding states. The $Q$-value spectra in (c) and (g) are described in the text.}
  \label{fig:Spectrum}
  \end{center}
  \vspace{-5mm}
\end{figure}

As aforementioned, $Q$-value resolution is of essential importance to differentiate various decay-paths in the present experiment. The reaction $Q$-value is defined as:
    \begin{equation}
        \label{eq1:3bodyQvalue}
        Q = E_{\rm {^2H}} + E_{\rm {^{\textit x}He}} + E_{\rm {^{\textit y}Be}} - E_{\rm beam}
    \end{equation}
where $\rm {^{\textit x}He}$ and $\rm {^{\textit y}Be}$ denote $\rm {^4He}$ + $\rm {^{12}Be}$ or $\rm {^6He}$ + $\rm {^{10}Be}$ decay pairs, and $E_{\rm beam}$ the beam energy. In most cases, only two outgoing particles are detected while the third one is deduced  by using the energy and momentum of the projectile~\cite{C16PRC2016, LIJING2017, C14CPC2018}. Due to the relatively large energy spread of the radioactive beam produced by projectile fragmentation (PF) type facility, the extracted $Q$-value spectra could hardly reach the required resolution~\cite{C16JPG2001, C16PRC2004, C16PRC2016, C14CPC2018}. To overcome this difficulty we directly measured all of the three final particles and deduced the beam energy event by event according to the energy-momentum conservation~\cite{LIJING2017}. Hence, the $Q$-value resolution lies solely on the performances of the detection system, but not on the beam energy uncertainty. Presently obtained $Q$-value spectra are shown in Figs.~\ref{fig:SpectrumQ1} and ~\ref{fig:SpectrumQ2} for the two final channels, respectively. For the first time, in PF-type experiments, $Q$-value peaks corresponding to the ground and low-lying excited states in the final fragments are clearly discriminated. For $^4$He decay channel (Fig.~\ref{fig:SpectrumQ1}),  the peak at about $-13.8$~\si{MeV} is for all three final particles in their ground states (ggg). Another peak at about $-15.9$~\si{MeV} is mainly associated with $\rm {^{12}Be}$ in its $2_1^+$ (2.109~\si{MeV}) state. The decay to the $0_2^+$ (2.251~\si{MeV}) state can not be resolved from this $Q$-value peak but would have much lower probability based on the analysis below. The decay to another nearby $1_1^-$ (2.715~\si{MeV}) state is less likely because it should stand at the far edge of the actual $Q$-value peak but apparently no structure appears there. For $\rm {^{6}He}$ decay channel (Fig.~\ref{fig:SpectrumQ2}), the highest peak at about $-16.5$~\si{MeV} is for the $Q_{\rm ggg}$, and another two at about $-19.8$~\si{MeV} and $-22.5$~\si{MeV} are associated with $\rm {^{10}Be}$ in its first excited state ($2_1^+$, 3.368~\si{MeV}) and the four adjacent states around $\sim$6~\si{MeV} ($2_2^+,~1_1^-,~0_2^+,~2_1^-$)~\cite{LIJING2017}, respectively.

The relative energies of $^{16}$C resonances can be derived from two breakup fragments using the standard invariant mass method~\cite{YANG2015, LIJING2017}. The excitation energy is the sum of the relative energy and the cluster separation threshold energy. Since the latter is related to the states of the final fragments, the excitation energy spectrum can be plotted by gating on a certain $Q$-value peak, corresponding to a selected decay path, as shown in Fig.~\ref{fig:Spectrum}.

The detection efficiencies for triple coincident events have been evaluated by Monte Carlo simulations (black-dashed lines in Fig.~\ref{fig:Spectrum}), taking into account of reaction kinematics, real experimental setup and detector performances. $^{16}$C is generated with a presumed exponential angular distribution, followed by an isotropic cluster-decay in the center of mass system~\cite{C16PRC2016, LIJING2017}. The relative energy resolution is simultaneously estimated, varying from 100 to 250~\si{keV} (FWHM) in the spectrum-covered ranges~\cite{YANG2015, LIJING2017}. The estimated production cross sections are about $3.25\pm0.19$~\si{mb} and $0.97\pm0.10$~\si{mb} for $\rm {^{4}He}$ and $\rm {^{6}He}$ channels, respectively, which are consistent with the previous reports~\cite{C16PRC2004}.

The excitation spectra in Fig.~\ref{fig:Spectrum} are fitted simultaneously by several resonant peaks (Breit-Wigner functions~\cite{BWPRL2007, BWPLB2012}), modified by detection efficiencies and convoluted with gaussian functions representing energy resolutions. The standard event-mixing background~\cite{YANG2014} has been evaluated but found to have negligible contributions to the spectra, possibly attributed to the rigorous timing matching of events as described above. The extracted resonances are listed in Table~\ref{tab::He4Ex}.

\begin{table}[htp!]
  \vspace{-1mm}
  \caption{
    \label{tab::He4Ex}
    Excitation energies, spin-parities and total decay widths of the resonances in $\rm {^{16}C}$, in comparison to those from the AMD calculations. Errors for positions and widths of the observed resonances are statistics only.
  }
  \resizebox{0.48\textwidth}{!}{
    \renewcommand{\arraystretch}{1.2}
    \begin{ruledtabular}
    \begin{tabular}{ccccc}

    \multicolumn{3}{c}{Present work}  &   \multicolumn{2}{c}{AMD calculations~\cite{C16Kimura2018}}\\
    \cline{1-3}
    \cline{4-5}
    $E_x$~\si{(MeV)} & $J^\pi$ & $\Gamma_{\rm tot}$~\si{(keV)} & $E_x$~\si{(MeV)} & $J^\pi$ \\
    \hline

    16.5(1)    &   $0^+$   &   1200(200)    &   16.81   &   $0_6^+$     \\
    17.3(2)    &           &   400(200)     &   17.51   &   $2_9^+$     \\
    18.3(1)    &           &   800(100)     &           &               \\
    19.4(1)    &           &   1500(160)    &   18.99   &   $4_{10}^+$  \\
    21.6(2)    &           &   2200(200)    &   21.49   &   $6_{5}^+$   \\
    23.5(2)    &           &   680(200)     &           &               \\
    25.5(2)    &           &   1230(200)    &           &               \\
    27.2(1)    &           &   1460(200)    &   29.30   &   $0_{14}^+$  \\
    \end{tabular}
    \end{ruledtabular}
  }
\end{table}

\begin{figure}
  \begin{center}
  \includegraphics[width=.435\textwidth]{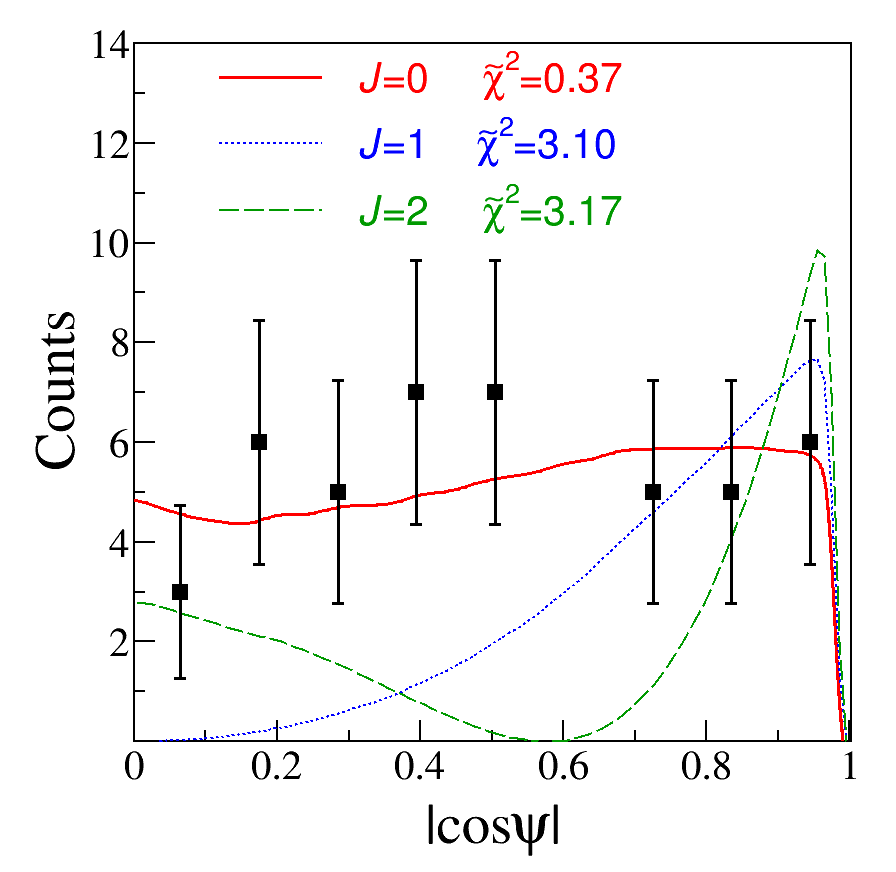}
  \caption{Angular correlation between $\rm {^4He}$ and $\rm {^{12}Be}$ fragments decaying from the 16.5~\si{MeV} resonance in $\rm {^{16}C}$. Experimental results are compared with theoretical distributions corrected by the detection efficiency. The corresponding reduced $\chi^2$ values are also presented accordingly.}
  \vspace{-7mm}
  \label{fig:Spin}
  \end{center}
\end{figure}

The latest AMD calculations~\cite{C16Kimura2018} have proposed a positive-parity linear-chain molecular band headed by the  16.81~\si{MeV} ${0^+}$ state which is close to the presently observed 16.5~\si{MeV} state (Fig.~\ref{fig:SpectrumE1}). Since little contamination was presented beneath this lowest energy peak, it would be adequate to apply the model-independent angular correlation analysis to determine its spin~\cite{SpinPRL1999, YANG2015, C16PRC2016, YangSpin2019}. For a spin-$J$ composite nucleus decaying into two spin-zero fragments, the projected angular correlation function can be formulated by a Legendre polynomial of order $J$, ${|{P_J}({\rm cos}[\psi+a{\theta^*}])|^2}$. Here $\psi$ is the polar angle of the relative velocity vector between the two fragments and ${\theta^*}$ the center-of-mass scattering angle of the resonant particle. $a$ is the phase shift correction factor which is not essential for small-angle scattering~\cite{SpinPRL1999} or $J=0$ resonances~\cite{YANG2015}. The presently obtained correlation function for 16.5~\si{MeV} state  (gated on $15.0\sim17.0$~\si{MeV}) is plotted in Fig.~\ref{fig:Spin} as a function of $|{\rm cos}(\psi)|$ owing to its symmetric feature about ${\rm cos}(\psi)=0$~\cite{YANG2015}. Experimental data are compared with the theoretical distributions assuming various $J$ values and corrected by the detection efficiencies, as demonstrated in Fig.~\ref{fig:Spin}. The best fit of the data is achieved with ${J^\pi}={0^+}$ whereas other spin assignments can be excluded due basically to the behavior at the minima and also to the much larger reduced $\chi^2$ values. We tried to use various cuts around the center of 16.5~\si{MeV} peak but no significant changes were found for the shape of the correlation spectra. Consequently, the observed 16.5 MeV state can be considered as the most promising candidate for the $0^+$ band head of the positive-parity linear-chain rotational band of $\rm {^{16}C}$. As a cross check, we tried also the standard angular correlation analysis \cite{C16PRC2016, YangPRC2019} for the observed 19.4~\si{MeV} state which is quite isolated in the channel decaying to $^{10}$Be(g.s.) (Fig.~\ref{fig:SpectrumE3}). It is found that, even though the low statistics do not allow a definite spin assignment, it is consistent with a spin-4 distribution. For other observed resonances, the spin determination would be impractical because of their overlaps with close-by states or the very low statistics.

\begin{figure}[htp!]
  \begin{center}
  \includegraphics[width=0.48\textwidth]{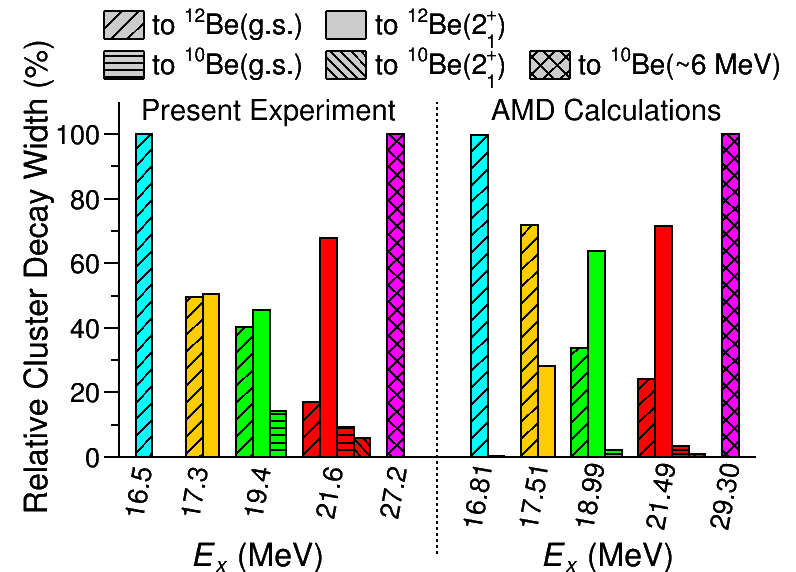}
  \caption{Relative cluster decay widths from the resonant states in $^{16}$C to the $\rm {^4He}$ + $\rm {^{12}Be}$ and $\rm {^6He}$ + $\rm {^{10}Be}$ final channels, extracted from presently observed spectra in Fig.~\ref{fig:Spectrum} (left panel) and AMD calculations (\cite{C16Kimura2018}) (right panel). The upward hatched, filled, plane hatched, downward hatched and cross hatched column bars represent decays to $^{12}$Be(g.s.), $\rm {^{12}Be({2_1^+})}$, $^{10}$Be(g.s.), $\rm {^{10}Be({2_1^+})}$ and $^{10}$Be($\sim$6~\si{MeV}) final states, respectively. Colors are used to differentiate the various excited states. Each partial cluster decay width is normalized to the sum of the widths from the the same mother resonance. }
  \vspace{-7mm}
  \label{fig:Branch}
  \end{center}
\end{figure}

As indicated qualitatively in some early works~\cite{vonOertzen2006, ClusterPPNP2015} and predicted quantitatively in recent AMD calculations~\cite{C16AMD2014, C14PRC2016, C16Kimura2018}, the decay from the mother resonance to certain states of the daughter fragments is closely related to the similarity of their structures. This structural link provides an important tool to probe the exotic structure in the former when a typical configuration has been clearly established in the latter~\cite{LIJING2017}. In the case of $^{16}$C, a positive-parity linear-chain molecular band, with the ${(3/2_\pi^-)^2}{(1/2_\sigma^-)^2}$ configuration, was predicted to have members at 16.81 ($0_6^+$), 17.51 ($2_9^+$), 18.99 ($4_{10}^+$), and 21.49 ($6_5^+$) MeV~\cite{C16Kimura2018}. Among them the $6^+$ member is predicted to possess peculiar decay features, as illustrated in Fig.~\ref{fig:Branch} (right panel). The large difference in partial decay width between its decays to the $\rm {^{12}Be({2_1^+})}$ and to the $\rm {^{12}Be({\rm g.s.})}$ states could partially be accounted for by the difference in penetration factors, but is still strongly related to the correlation between the chain-like structure in $^{16}$C(21.6 MeV) and the angular momentum in the daughter nucleus ~\cite{C16Kimura2018, ClusterPPNP2015}. From the experimental side, the observed 21.6 MeV state is close to the predicted 21.49 MeV ($6_5^+$) state (Table~\ref{tab::He4Ex}). Adopting a spin-parity of $6^+$, this state should decay with higher probability to the $\rm {^{12}Be({2_1^+})}$ state than to the $\rm {^{12}Be({0_2^+})}$ state, because of the more than five times larger penetration factor for the former than for the latter. This observed decay is also much stronger than that to the $\rm {^{12}Be(g.s.)}$, being well consistent with the prediction. The theoretical calculations also predict small partial decay widths for the ground and first excited states of $^{10}$Be, which are also perfectly confirmed by our experimental observations, as displayed in Fig.~\ref{fig:Spectrum} and plotted quantitatively in Fig.~\ref{fig:Branch} (left panel) for the 21.6 MeV state. As a consequence, the observed 21.6 MeV resonance should be regarded as the $6^+$ member of the predicted positive-parity linear-chain molecular band of $\rm {^{16}C}$, despite the lack of direct spin measurement. We also assign the $2^+$ and $4^+$ members of the band to the observed resonances at 17.3 and 19.4 MeV states, considering their similarities in excitation energies and selective decay properties (Fig.~\ref{fig:Branch}). The observed 18.3~\si{MeV} state  (Fig.~\ref{fig:Spectrum} and Table~\ref{tab::He4Ex}) is also quite close to the proposed $4^+$ member but actually not classified into the present positive-parity band due to its primary decay path to $^{12}$Be(g.s.) and negligible decay to the $^{10}$Be channel, which is contradictory to the prediction. This additional state with a quite large $\alpha$-decay probability might belong to other molecular configurations~\cite{C16Kimura2018}. As for the 16.5~\si{MeV} state, the above spin-zero band-head assignment can be further confirmed by its pure decay to $\rm {^{12}Be}$(g.s.), in agreement with the theoretical prediction. We note that the systematic error for these relative decay widths is estimated to be less than 5$\%$, due basically to uncertainties in simulation and detection.

It is worth noting that the previously reported peak at about 20.6~\si{MeV}, reconstructed from the $\rm {^{6}He}$ + $\rm {^{10}Be}$ channel without $Q$-value selection~\cite{C16PRC2016}, is not observed in our measurement. This prior peak might be understood by erroneously shifting the presently observed 23.5~\si{MeV} peak in Fig.~\ref{fig:SpectrumE4} and 27.2~\si{MeV} peak in Fig.~\ref{fig:SpectrumE5} into Fig.~\ref{fig:SpectrumE3}, according to their different $Q$-values.

Another intriguing high-lying state at 27.2~\si{MeV}  (Fig.~\ref{fig:SpectrumE5} and Table~\ref{tab::He4Ex}) is found to decay primarily into the $\sim6$~\si{MeV} states in $\rm {^{10}Be}$. We have made further investigations with AMD method to explain the states in $^{16}$C at very high excitation energies, where a novel linear-chain molecular band with ${(1/2_\sigma^-)^2}{(1/2_\sigma^+)^2}$ configuration appears, which decays predominantly to the $0_{2}^+$ (6.179~\si{MeV}) state of $\rm ^{10}Be$. The property of the presently observed 27.2~\si{MeV} state in  $^{16}$C (Fig.~\ref{fig:Branch}) agrees quite well with the predicted band head state ($0_{14}^+$). Further experimental investigations are certainly encouraged to clarify the existence of this very high-lying linear-chain molecular band in $^{16}$C.

In summary, a new inelastic excitation and cluster-decay experiment was carried out for $\rm ^{16}C$ and the triple coincident detection with quite high efficiency was realized.  For the first time, in PF-type measurements, good $Q$-value resolution was achieved for both $\rm {^4He}\!+\!{^{12}Be}+\!{^{2}H}$ and $\rm {^6He}\!+\!{^{10}Be+\!{^{2}H}}$ final channels, allowing the reconstruction of $^{16}$C resonances according to their decay paths. The systematic decay-pattern analysis and the spin determination for the band head fully support the existence the ${(3/2_\pi^-)^2}{(1/2_\sigma^-)^2}$-type linear-chain molecular band in $^{16}$C, as predicted by the latest AMD calculations \cite{C16AMD2014,C16Kimura2018} and by the earlier molecular-orbital approach \cite{C16MOcal2001}.  Moreover, an exotic high-lying excited state at 27.2~\si{MeV} is found to decay dominantly to the $\rm {^{10}Be(\sim6~\si{MeV})}$ state, in line with the predicted ${0^+}$ member of ${(1/2_\sigma^-)^2}{(1/2_\sigma^+)^2}$ linear-chain molecular band at even higher energies. It would be very interesting to further investigate the clustering structures in $^{16}$C at even higher excitation domain where the pure $\sigma$-bond and the high-lying negative-parity molecular bands may be accommodated.
\begin{acknowledgments}
The authors wish to thank the staff of HIRFL-RIBLL for their technical and operational support.
This work was supported by the National Key R$\&$D Program of China (Grant No. 2018YFA0404403) and the National Natural Science Foundation of China (Grant Nos. 11535004, 11875073, 11875074, 11961141003, 11775004, 11775003).
\end{acknowledgments}

\bibliography{ref-16C}
\end{document}